\documentclass[12pt]{article}
\usepackage{a4,epsfig,rotating,amssymb}
\begin{document}
\vspace*{1cm}

\begin{center}
\begin{Large}
\boldmath
\bf Development of scintillating fiber detector technology
  for high rate particle tracking 
\unboldmath 
\end{Large}
\vspace*{1cm}
\begin{Large}
E.C.Aschenauer, J.B\"ahr, V.Gapienko\footnote{on leave from
  IHEP Protvino, Russia} , B.Hoffmann\footnote{now at Esser Networks
  GmbH, Berlin},
A.Kharchilava\footnote{on leave from the Institute of Physics, Georgian
  Academy of Sciences, Tbilisi},
H.L\"udecke, R.Nahnhauer, R.Shanidze\footnote{on leave from High
  Energy Physics Institute, Tbilisi State University} \\

\vspace*{0.5cm}
DESY-IfH Zeuthen\\
\vspace*{1cm}
\end{Large}
\end{center}



\begin{abstract}

The performance of a scintillating fiber detector prototype for tracking under
high rate conditions is investigated. A spatial resolution of about
100\,$\mu$m is aimed for the detector. Further demands
are low occupancy and radiation hardness up to 1\,Mrad/year.
Fibers with different radii and different wavelengths of the
scintillation
 light from different producers have been extensively tested
concerning light output, attenuation length and radiation hardness,
with and without coupling them to light guides of different length and
diameter.

In a testrun at a 3 GeV electron beam the space dependent efficiency
and spatial resolution of fiber bundels were measured by means of two
external reference detectors with a precision of 50\,$\mu$m.
The light output profile across fiber roads has been determined
with the same accuracy.

Different technologies were adopted for the construction of tracker
modules consisting of 14 layers of 0.5\,mm fibers and 0.7\,mm pitch.
A winding technology
provides reliable results  to produce later fiber modules of about 
25$\times$25\,cm$^2$ area.

We conclude that on the basis of these results a fiber
tracker for high rate conditions can be built.

\end{abstract}

\vspace*{2cm}

Contribution to the International Europhysics Conference on High Energy
Physics , August 1997, Jerusalem

\newpage 
 
\section { Introduction}
\renewcommand{\floatpagefraction}{0.95}

The use of scintillating fiber detectors has some advantages compared
to other detector principles in terms of spatial and time resolution,
robustness of the detector, match to different shapes, radiation hardness,
etc. \cite{rn}. Examples for the advantageous use of fiber detectors
under very different conditions are the D0
experiment, CHORUS \cite{CHOR} and the H1 Forward Proton
Spectrometer \cite{WARS}.

The fiber detector under discussion is aimed to be a tracking device,
with time characteristics according to the bunch crossing time
of the accelerator of 96\,ns. 
The spatial
resolution is required to be about 100\,$\mu$m. The fiber detector should be
of such a granularity, that an occupancy of a few percent is reached.
The structure and readout of the detector has to be constructed in
such a way, that up
to four events per bunch crossing can be registered with an overall charged
particles multiplicity of more than 100. 
The scintillating fibers should not change their characteristics
significantly after an irradiation of 1$\div$2\,Mrad.
All these demands and the solutions presented below match to a
possible application of the fiber detector as the inner tracker in the
HERA-B project at DESY \cite{HERAB}.
The fiber detector would consist of 48 one-dimensional detector
planes, some of them operating in a magnetic field of about
0.8\,T.
A plane consists of four quadrants of 25$\times$25\,cm$^2$ each.
The available space and the magnetic field conditions demand the light
collection from
the scintillating fibers by means of light guides of a length of about
2$\div$3\,m.
The readout of the scintillating fiber detector is assumed to be
realized by multichannel photomultipliers (PSPM) of
Hamamatsu\footnote{Hamamatsu Photonics K.K., Electron Tube Division,
  314-5, Shimokanzo, Toyooka Village, Iwata-gun, Shizuoka-ken, Japan}
type R5900-00-M64 with 64 pixels per device. This photomultiplier is
still under development, only a few prototype examples exist. The
characteristics are similar to the 16 pixel devices used for our
investigations.
The test and characteristics of this PSPM are not more
subject of this paper.


In chapter 2 the optical properties (light yield, attenuation length
and coupling efficiency) and radiation hardness of various fiber
materials are dicussed. \\
In chapter 3 the results obtained for fiber detector prototypes in a test run at a 3 GeV electron-beam are
presented. For different fiber bundels efficiency and spatial
resolution
were determined by means of reference
detectors.\\
The development of the technology for the large scale production of
fiber detectors is described in chapter 4.

\section {Choice of fiber material}

\subsection {Optical characteristics}

\subsubsection*{Method}

All measurements were performed with standardized fiber samples of
30\,cm length and a cross section of 2$\times$2\,mm$^2$ independent of the fiber
diameter, which varies between 0.25\,mm and 0.50\,mm. Fibers of three
producers (BICRON\footnote{BICRON, 12345 Kinsman Road, Newbury,
  Ohio, USA},
 KURARAY\footnote{KURARAY Co. LTD., Nikonbashi, Chuo-ku, Tokyo 103, Japan},
Pol.Hi.Tech.\footnote{Pol.Hi.Tech., s.r.l., S.P.Turanense, 67061
  Carsoli(AQ), Italy}) were investigated,
whereby the wavelength of the emitted scintillation ligth covered the
blue and green spectral regions. All investigated fibers have a double
cladding, which leads to an increased light trapping efficiency.
The sample was connected to two Philips\footnote{Philips Photonique,
  Av. Roger Roncier, B.P.520, 19106 Brive, France}
photomultipliers (PMs) XP2020, S1 and
S2. Below the fiber sample a scintillator (5\,mm thick and 10\,mm width)
was installed. It was readout by two PMs Philips XP1911, T1 and T2, from each
side. The sample was exposed to a $^{106} Ru$ source.
A collimator with variable slit width was mounted between
source and fiber sample. The amplitude spectra were measured by an ADC,
if a trigger signal occured, derived from a coincidence between S2, T1
and T2. The setup was calibrated so that the number of
photo-electrons (pe) could be estimated. The results are related to the
bialkaline photo-cathod of the PM XP2020 which is similar to the bialkaline
photo-cathode of the multi-channel PM forseen for later application.

\subsubsection*{Light yield}

\begin{figure}[ht]
  \setlength{\unitlength}{1cm}
  \begin{picture}(12.0,10.0)
     \put(1.0,-1.0){\epsfig{file=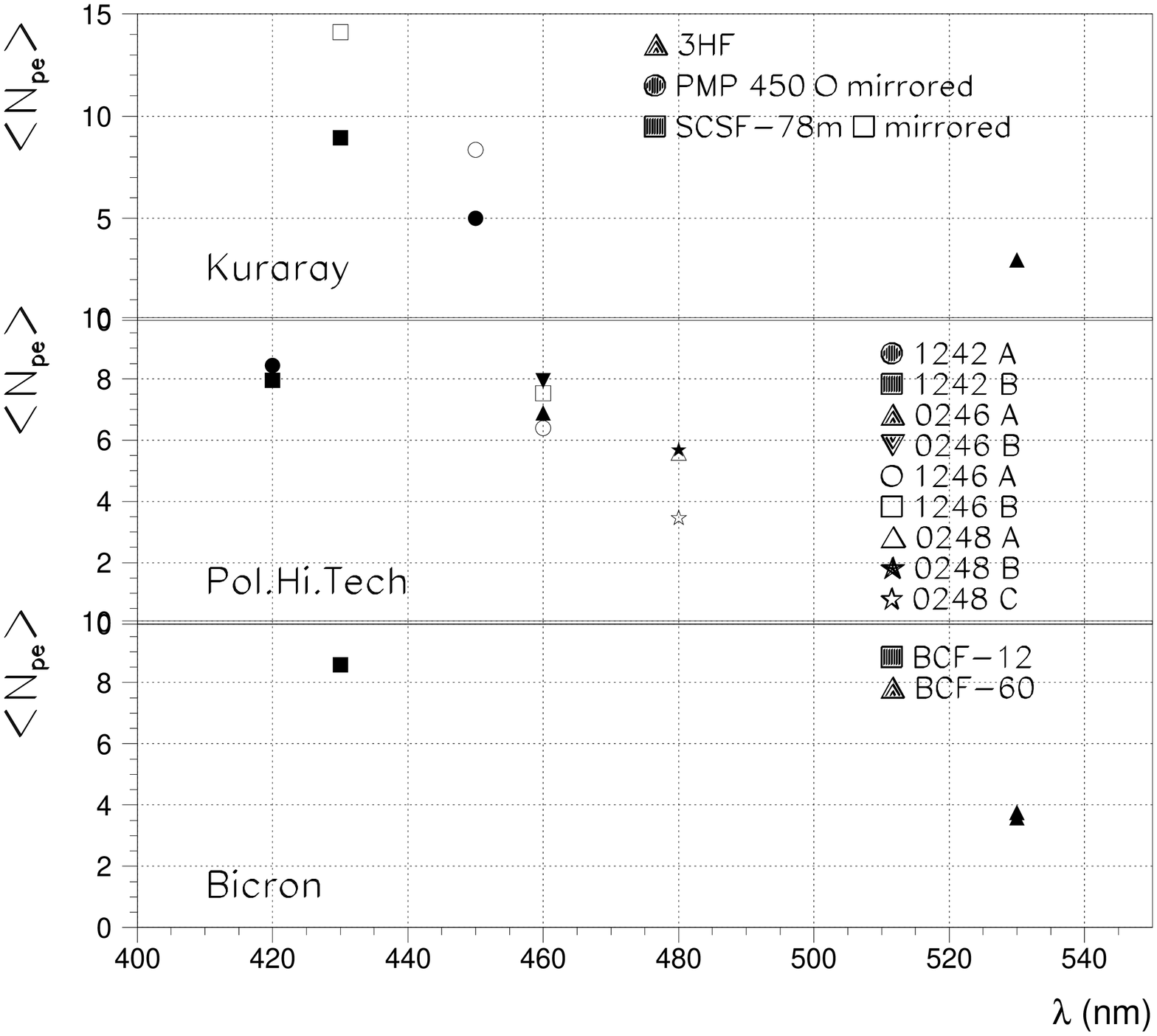,width=12cm}}
  \end{picture}
  \caption{Light output for different scintillation fiber materials (diameter:
    0.5\,mm) from the companies KURARAY, Pol.Hi.Tech. and BICRON. }
  \label{fig3}
\end{figure}

The results for a sub-sample of fiber materials of 0.5\,mm diameter
are shown in Fig.\ref{fig3}. Generally, it is seen that the light yield
decreases with increasing scintillator emission wavelength because the PM's
sensitivity curve is not unfolded. There is no remarkable difference 
between the best materials of the three producers and the light yield
is typically
4.5\,pe per 1\,mm scintillator.
\begin{center}
\begin{tabular}{|l|l|}\hline
        Producer & Material\\ \hline  
	BICRON	& BCF 12\\
	KURARAY	& SCSF-78M\\
	Pol.Hi.Tech & POLIFI 1242A and B\\ \hline
\end{tabular}
\end{center}

The application of a mirror on one side of the fiber sample
increases the light yield by a factor of 1.7. The light yield decreases
with decreasing fiber diameter by 10$\div$40 percent for diameters of
0.5\,mm compared to 0.25\,mm.

\subsubsection*{Attenuation in clear fibers}

\begin{figure}[h]
  \setlength{\unitlength}{1cm}
  \begin{picture}(12.0,13.0)
     \put(2.0,-1.0){\epsfig{file=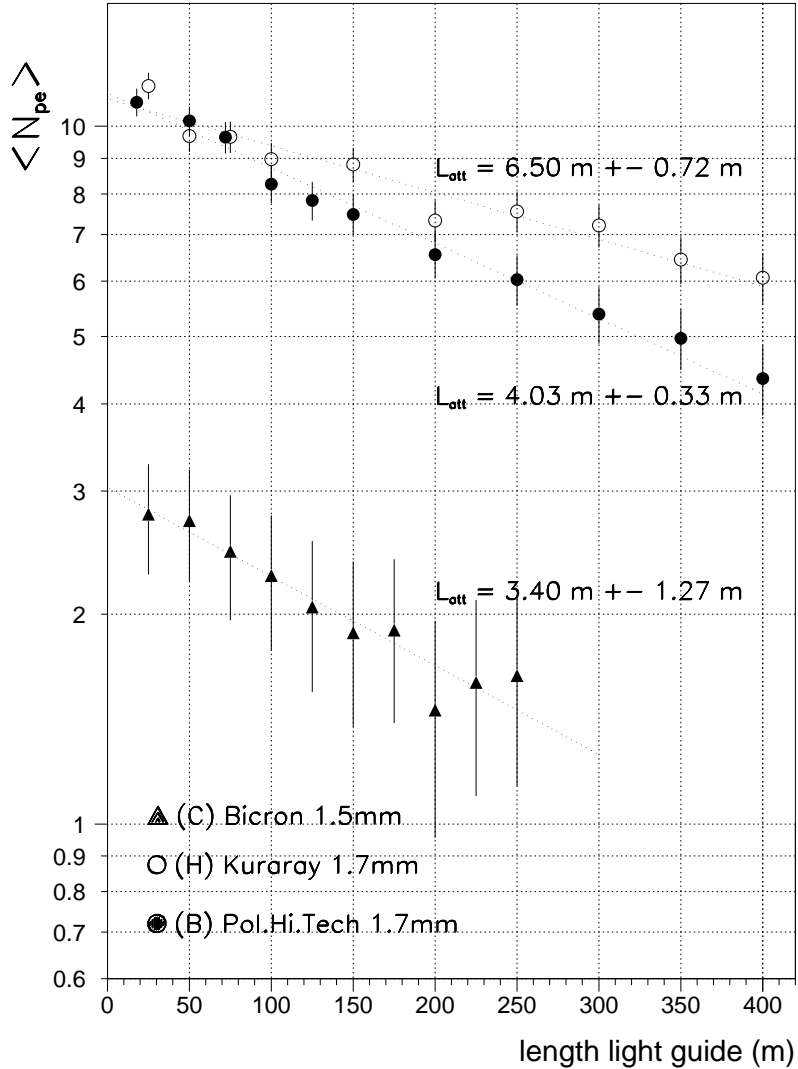,width=12cm}}
  \end{picture}
  \caption{Attenuation length for clear fibers (diameters: 1.0\,mm, 
   1.5\,mm, 1.7\,mm) produced by KURARAY, Pol.Hi.Tech. and BICRON. }
  \label{fig5}
\end{figure}

The measurement of the attenuation length of clear fibers is done with
the setup previously described. A scintillating fiber sample of known light yield is used. The
clear fiber is coupled to the fiber sample  by a
standardized coupling mask. The measurements were performed for
clear fibers of different diameters (1.0\,mm...1.7\,mm)
from the three producers. 
The results are shown in Fig.\ref{fig5}. 
No strong dependence on fiber
diameter and wavelength is seen. The fibers
from KURARAY show the best attenuation length for all diameters.

\subsubsection*{Coupling of scintillating fibers and clear fibers}

Several tests were performed to couple scintillating and light guide
fibers. After an optimization of the coupling pieces the coupling efficiency
became better then 95\%  independent on the medium between both fibers (air, glue,
optical grease).

\subsection{ Radiation hardness}

Radiation hardness tests were performed depositing doses of
about 1\,Mrad on scintillating fiber samples in a few minutes. The
fiber samples are similar to those described above (see chapter 2.1).

Irradiations were performed in a 70\,MeV energy proton beam
at the VICKSY accelerator of the Hahn-Meitner Institute, Berlin. The
beam leaves the beam tube through a scatter foil, passes a 380\,mm
air gap
and two diaphragms of 30\,mm and 50\,mm thick PMMA shaping a radiation
field of 2\,mm$\times$10\,mm on the fiber sample. The accumulated dose
is measured by an
ionization chamber behind the sample. 

Different scintillating fiber materials of the three producers mentioned
above were irradiated. The influence of using glue in the sample
production on the radiation hardness was also studied.The
scintillating fiber samples were
irradiated as follows:

\begin{itemize}
\item Spot-like irradiation.\\
 The scintillating fiber samples were irradiated at two
  places along the fiber, 10\,cm from the fiber ends. The
  accumulated doses have been 1$\div$1.4\,Mrad and 0.2$\div$0.4\,Mrad
  at the two positions, respectively.

\item Profile-like irradiation.\\
 Along the scintillating fiber sample the accumulated radiation dose
 was decreased from 1\,Mrad to 0.2\,Mrad.

\end {itemize}

The light output was measured 
before and after the irradiation (for several weeks)
at different points of the irradiated sample so that the influence of
high and low dose could be distinguished. Also the measurement positions
are chosen such, that the damage of scintillator efficiency 
and/or plastic matrix
(light transmission) can be disentangled.

\subsubsection*{Results}

The results are given in table \ref{table4}. For most of the materials
the scintillator efficiency decreases by 20$\div$80 percent just
after the irradiation; the decrease in light
transmission varied from 20 to 70 percent compared to the initial value.

\begin{figure}[h]
  \setlength{\unitlength}{1cm}
  \begin{picture}(10.0,12.0)
     \put(2.0,-1.0){\epsfig{file=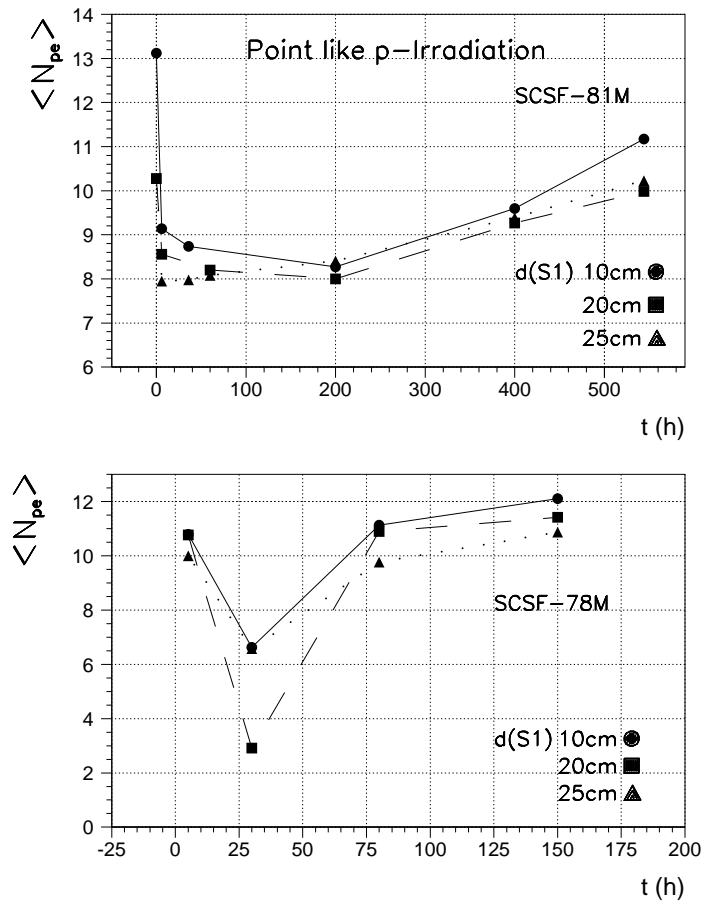,width=10cm}}
  \end{picture}
  \caption{Evolution in time of the light output for point-like irradiated
    KURARAY SCSF scintillating fibers. The solid, dashed and dotted curves
    correspond to measurements with the source placed at 10, 20 and
    25\,cm with respect to S1, respectively.}
  \label{fig.10}
\end{figure}

For four materials from Pol.Hi.Tech. only transmission is damaged.For
nearly all materials a strong recovery process is seen. It takes from
80 to 600 hours to recover the light yield and transmission to a level
of at least 90\%. No significant influence of the glue on damage
and recovery is observed. Fig.\ref{fig.10} shows the behaviour of two
samples for several hundred hours after irradiation.

\section {Results of test run in an electron-beam}


Small-scale fiber detector prototypes were exposed in an 3\,GeV e-beam
at DESY in
order to measure:

\begin{itemize}

\item the efficiency and resolution

\item the light output across the fibers.

\end {itemize}

Here, a detector geometry is defined which is the basis for all further
investigations. The fiber detectors are assumed to be constructed of 14
layers of 0.5\,mm scintillating fibers. The fibers are arranged with a
pitch of 700\,$\mu$m  in the layers. The layers are staggered to each other
by 350\,$\mu$m. The seven scintillating fibers with the same
coordinate form a "road " and are coupled to one light guide fiber of
1.7\,mm diameter.

\begin{figure}[h]
  \setlength{\unitlength}{1cm}
  \begin{picture}(13.0,5.0)
     \put(3.0,-4.5){\epsfig{file=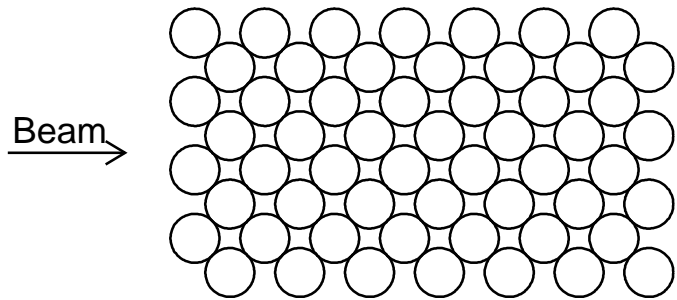,width=13cm}}
  \end{picture}
  \caption{Schematic cross section of the exposed fiber bundle}
  \label{crossec}
\end{figure}

The fiber samples used in the test run are based on this geometry
defined for the final detector. Fig.\ref{crossec} shows a schematic
cross section
through the fiber bundle exposed in the test beam. It consists of 8
roads with 7 fibers per road. The
diameter of fibers is
0.5\,mm. The nominal pitch in one layer amounts to 700\,$\mu$m.

\begin{figure}[h]
  \setlength{\unitlength}{1cm}
  \begin{picture}(14.0,8.0)
     \put(1.0,0.0){\epsfig{file=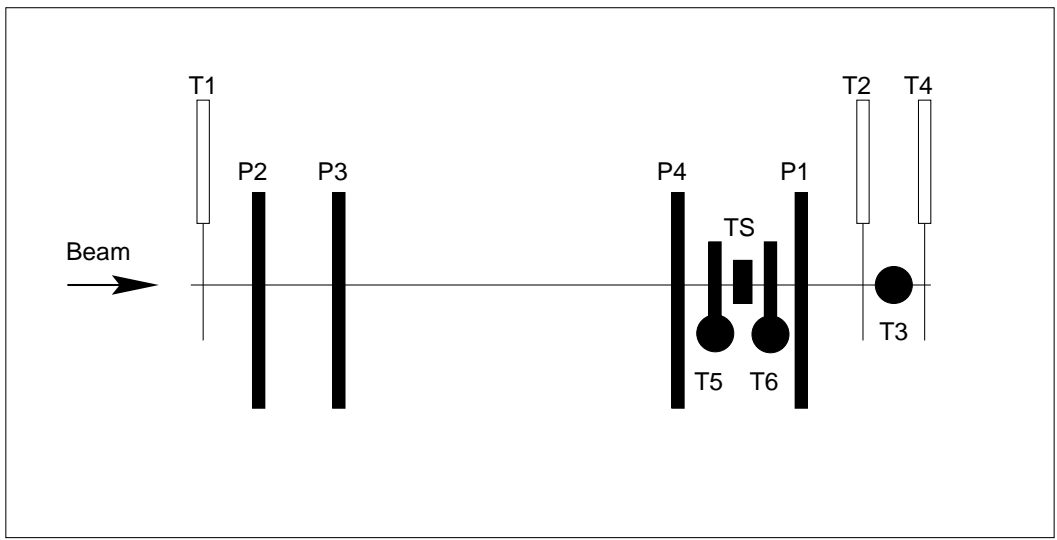,width=14cm}}
  \end{picture}
  \caption{Setup of the test-beam exposure; T1...T4: Trigger counters,
    T5...T6: Planes of the fiber reference detector included in the trigger,
    P1...P4: Planes of the Si-strip telescope, TS: Fiber sample
    under test}
  \label{setup}
\end{figure}

\begin{figure}[ht]
  \setlength{\unitlength}{1cm}
  \begin{picture}(10.0,9.0)
     \put(2.5,-1.0){\epsfig{file=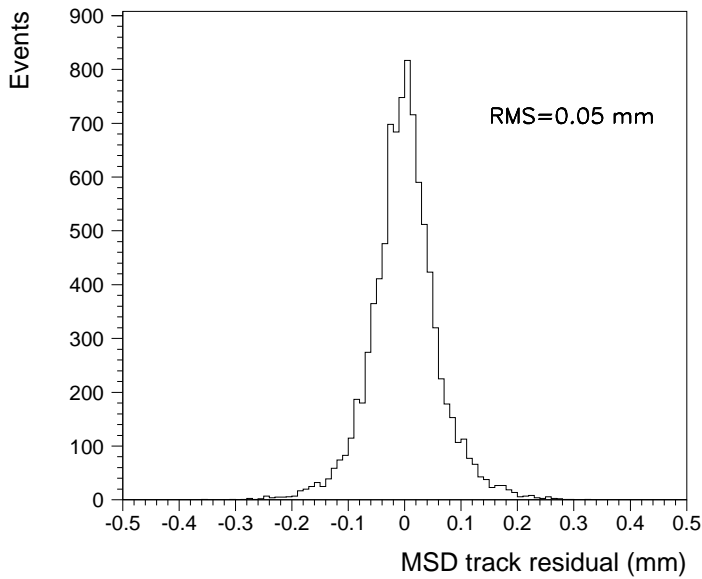,width=10cm}}
  \end{picture}
  \caption{Track residual of the Si-strip telescope}
  \label{MSD}
\end{figure}

The setup of the beam tests is scetched in fig.\ref{setup}. A similar
setup was described in more detail in \cite{setup}. It consists of a
trigger system, two external reference detectors and
the fiber sample itself. The scintillation light is collected via 3\,m
long light guide fibers. The reference detectors are a scintillating
fiber detector consisting of two planes (T5,T6) giving an accuracy of
170\,$\mu$m for
through-going tracks and a Silicon micro-strip telescope (P1$\div$P4).
For the
Si-telescope a track residual of 52\,$\mu$m  was measured as shown in
Fig.\ref{MSD}.

\subsubsection* {Results}

The mean light output for roads of 7 fibers readout via 3\,m long light
guide fibers was measured to be 6.2 photo-electrons. 

\begin{figure}[ht]
  \setlength{\unitlength}{1cm}
  \begin{picture}(10.0,9.0)
     \put(2.5,-1.0){\epsfig{file=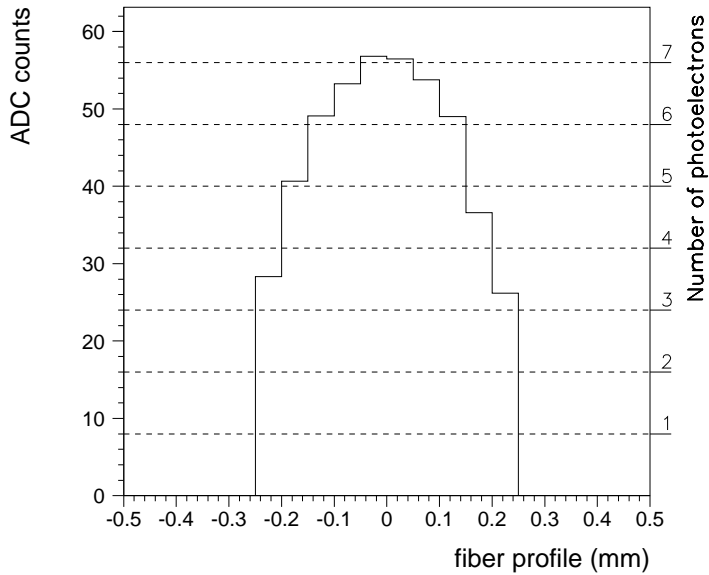,width=10cm}}
  \end{picture}
  \caption{Light output across the fiber road}

  \label{profile}
\end{figure}

The "light profile", i.e. the light output across the fiber road
is shown in Fig.\ref{profile}.
It follows the expected dE/dx behaviour for the realized fiber
geometry.

The efficiency for all roads plotted in Fig.\ref{effic}  shows a flat
distribution with a mean value of 98 percent.

\begin{figure}[ht]
  \setlength{\unitlength}{1cm}
  \begin{picture}(9.0,9.0)
     \put(2.0,-1.0){\epsfig{file=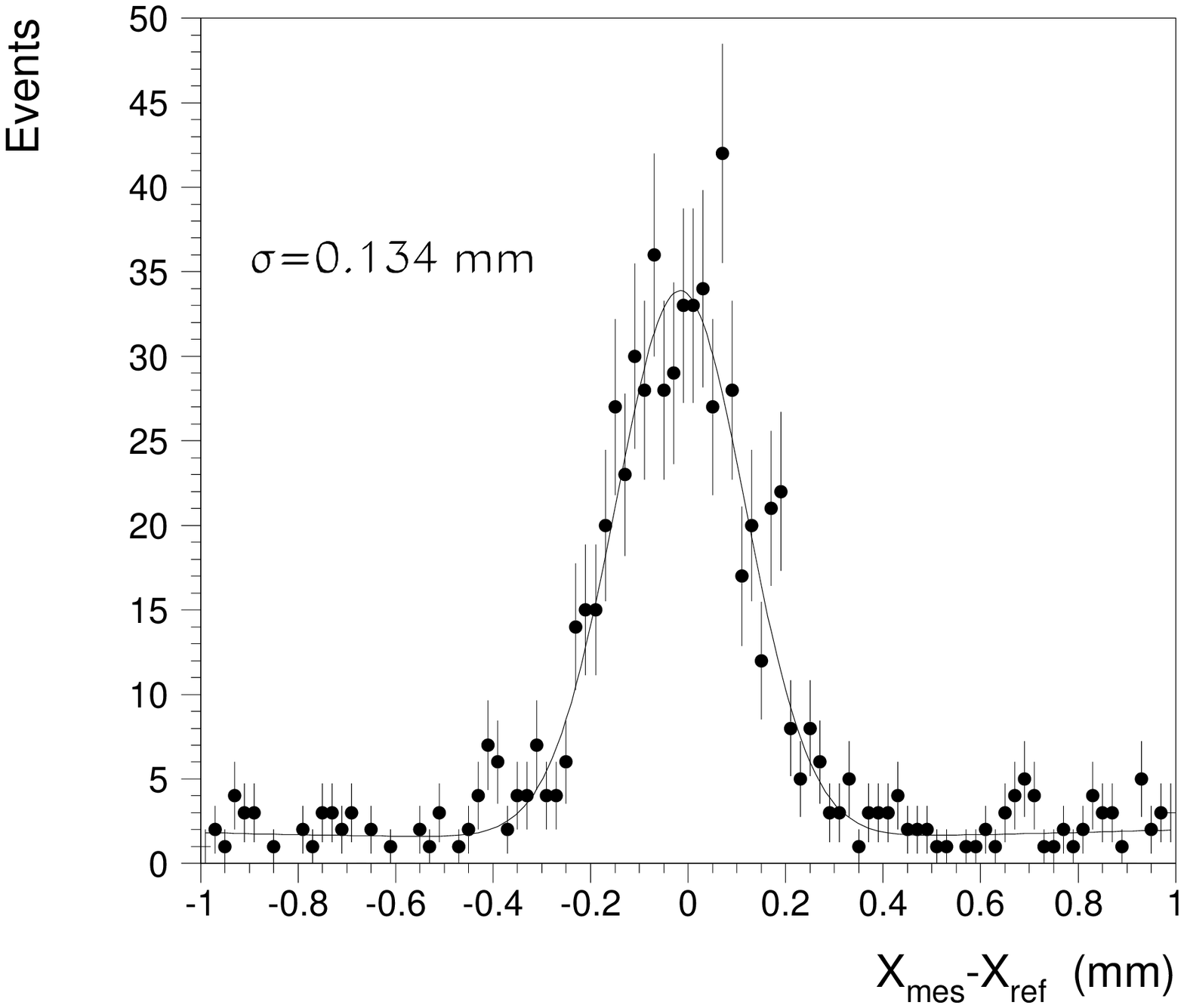,width=10cm}}
  \end{picture}
  \caption{Spatial resolution of the fiber sample}

  \label{resol}
\end{figure}

\begin{figure}[ht]
  \setlength{\unitlength}{1cm}
  \begin{picture}(9.0,9.0)
     \put(2.0,-1.0){\epsfig{file=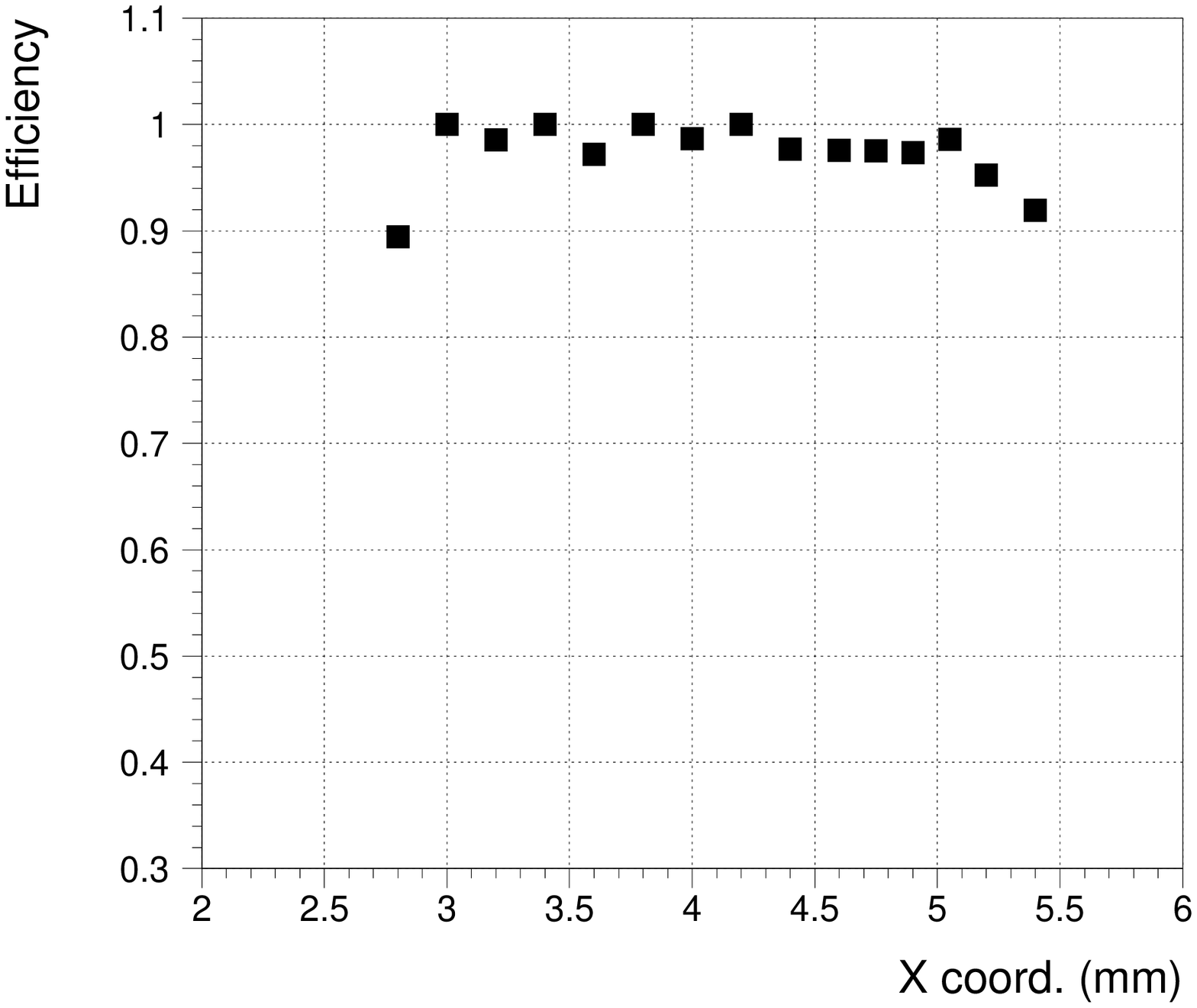,width=10cm}}
  \end{picture}
  \caption{Efficiency distribution of the fiber sample}

  \label{effic}
\end{figure}

The spatial resolution was measured to be 121\,$\mu$m
(Fig.\ref{resol}) by the reference
detectors unfolding the accuracy
of 52\,$\mu$m of the Si-telescope.

\section{ Development of fiber detector technology}

Different technologies are tested
to find the best way to produce the fiber detector modules:

\begin{itemize} 
\item A winding technology as it was used for the production of the
  CHORUS tracker \cite{CHO}. 

\item A mounting technology of single layers based on a
 proposal from the Heidelberg
 University \cite{HD}. A prototype is produced by
 GMS\footnote{GMS - Gesellschaft f\"ur Me\ss-und
  Systemtechnik mbH, Rudower Chaussee 5, 12489 Berlin, Germany}.

\item A technology developed by KURARAY were also a
 prototype is produced.
\end{itemize}

Using these methods several prototypes are partially still under construction
and could not be investigated up to now in detail. 
 This will be done however in a forthcoming testrun at DESY.


The winding technology was investigated in more detail in our lab. A
construction setup was manufactured and tested.
The principle of the winding technology is based on a layer-by-layer
increase of the
mechanical tension of the fiber to receive a flat detector
after removing the ribbon from the winding drum.
The results are encouraging; flat ribbons with a good accuracy can be
produced.



\section{ Conclusions}

Based on the investigations presented above we conclude,
that a radiation hard fiber detector for high rate conditions
fulfilling 
the demands on efficiency, time and spatial
resolutions can be realized.

Extensive investigations resulted in a material choice, which gives
a light yield of 4.5\,pe/mm  and
radiation hardness of at least 1\,Mrad for the material SCSF-78M.

In the exposure of test detector samples to an electron  beam  a spatial
resolution of 121\,$\mu$m is measured. The efficiency is
constant across the detector and amounts to
 about 98\%. 

The time delay and jitter resulting from the effects of light
collection in fibers and readout via multi-channel
PMs are of the order of a few nanoseconds not considering readout
electronics.

Different technologies for the production of the fiber modules are
tested. The comparison of the technologies is still going on. We
conclude, that the
construction of fiber detectors  for high rate conditions on the basis
of these investigations seems to be possible.

\section*{Acknowledgement}

Part of this work was done in close collaboration with groups from the
universities of Heidelberg and Siegen. We want to thank our colleagues
for their good cooperation and many fruitful discussions.

The fiber irradiation tests were possible only due to the kind support 
of the Hahn-Meitner-Institute Berlin. We are deeply indebted to the
VICKSY accelerator team and want to thank in particular
Dr.D.Fink, Dr.K.Maier and Dr.M.Mueller from HMI and Prof.Klose from GMS for a 
lot of practical help.

We acknowledge the benefit from the DESY II accelerator crew and
the test area maintainance group.



\newpage
\renewcommand{\arraystretch}{1.2}
\begin{sidewaystable}
{\small
\begin{tabular}{|c|c|c|c|c|c|c|c|} \hline \hline
{\bf Material} & {\bf $\lambda$(SF)} & {\bf Specialties} & {\bf Irradiation} & 
{\bf Dose at } & {\bf Damage (\%) at } & {\bf Recovery to 90 \% } & {\bf Result} \\
   &      &      &   & {\bf 10 - 20 cm} & 
\multicolumn{2}{|c|}{ \bf 10 - 20 - 25 cm} & \\ \hline
BCF-12   & 430 & glue & spot & 0.4 / 1.4 Mrad & 72 - 82 - 60 & 600 h & T and S damaged \\ \hline
BCF-12   & 430 & no glue & spot & 0.4 / 1.4 Mrad & 62 - 62 - 68 & 600 h & T and S damaged \\ \hline
BCF-60   & 530 & glue & spot & 0.4 / 1.4 Mrad & 58 - 76 - 60 & 600 h & T and S damaged \\ \hline
BCF-60   & 530 & no glue & spot & 0.4 / 1.4 Mrad & 63 - 45 - 49 & $>$ 600 h & T and S damaged \\ \hline
BCF-12   & 430 & glue & profile & 0.2 - 1.0 Mrad & 82 - 77 - 82 & 160 h & T and S damaged \\ \hline
BCF-60   & 530 & glue & profile & 0.2 - 1.0 Mrad & 95 - 89 - 89 & 160 h & T and S slightly damaged \\ \hline
\hline
0042-2-0975 & 430 & glue & spot & 0.4 / 1.4 Mrad & 79 - 89 - 80 & 400 h & T and S slightly damaged \\ \hline
0042-2-0975 & 430 & no glue & spot & 0.4 / 1.4 Mrad & 70 - 100 - 85 & 400 h & T and S slightly damaged \\ 
\hline
1242 A   & 420 & glue & spot & 0.2 / 1.4 Mrad & 99 - 73 - 62 & $>$ 180 h & mainly T damaged \\ \hline
0246 B   & 460 & glue & spot & 0.2 / 1.4 Mrad & 100 - 99 - 94 & 0 h & no damage at all \\ \hline
0248 C   & 480 & glue & spot & 0.2 / 1.4 Mrad & 100 - 93 - 89 & 150 h & mainly T damaged \\ \hline
1242 B   & 420 & glue & profile & 0.2 - 1.2 Mrad & 100 - 52 - 27 & $>$ 160 h 100 - 48 - 47 & mainly T damaged \\ \hline
1246 B   & 460 & glue & profile & 0.2 - 1.2 Mrad & 100 - 99 - 83 & 160 h & mainly T damaged \\ \hline 
\hline
SCSF-81M & 430 & glue & spot & 0.4 / 1.4 Mrad & 70 - 83 - 77 & 600 h & T and S damaged \\ \hline
SCSF-81M & 430 & no glue & spot & 0.4 / 1.4 Mrad & 80 - 78 - 100 & 400 h & S damaged \\ \hline
SCSF-78M & 430 & glue & spot & 0.2 / 1.4 Mrad & 72 - 33 - 74 & 80 h & T and S damaged \\ \hline
PMP-450  & 450 & glue & spot & 0.2 / 1.4 Mrad & 100 - 100 - 100 & 0 h & no damage at all \\ \hline
3HF      & 530 & glue & spot & 0.2 / 1.4 Mrad & 100 - 100 - 100 & 0 h & no damage at all \\ \hline
SCSF-78M & 430 & glue & profile & 0.2 - 1.2 Mrad & 94 - 89 - 84 & 180 h & T and S damaged \\ \hline
PMP-450  & 450 & glue & profile & 0.2 - 1.2 Mrad & 95 - 79 - 76 & 180 h & T and S damaged \\ \hline
3HF      & 530 & glue & profile & 0.2 - 1.2 Mrad & 100 - 100 - 100 & 0 h & no damage at all \\ \hline
\hline
%
\end{tabular}}

\caption{\label{table4} Results on the proton irradiation damage
to the different scintillating fiber materials.} 
 
\end{sidewaystable}


\bigskip
\vspace{2cm}

\begin{thebibliography}{100}


\bibitem{rn}R.Nahnhauer ed., Application of Scintillating Fibers in
Particle Physics, Proceedings of Workshop, Blossin, GDR, 1990

\bibitem{CHOR}N.Armenise et al., CERN-SPSC-90-42, Geneva


\bibitem{WARS}H1 Collaboration, The Forward Proton Spectrometer of
H1, 28th International Conference on High Energy Physics ICHEP'96,
Warsaw, Poland

\bibitem{HERAB}T.Lohse et al., HERA-B Technical Proposal, DESY-PRC 94/02 (1994)
 
\bibitem{setup}J.B\"ahr et al., Nucl. Instr. and Meth. A371 (1996)
380-387


\bibitem{CHO}T.Nakano et al.,  Proceedings of the Workshop on
Scintillator Fiber Detectors (ed.A.D.Bross et al.), SCIFI 93,
Notre Dames, USA, 1993, p.525
 
\bibitem{HD}F.Eisele et al., Private communication
 

\end{thebibliography}
\end{document}